\documentclass[journal=jacsat,manuscript=article]{achemso}

\usepackage[version=3]{mhchem} 
\usepackage{siunitx}
\usepackage{upgreek}

\newcommand{\stxiv}[2]{#2} 

\author{Michal Horák}
\affiliation[CEITEC]
{Brno University of Technology, Central European Institute of Technology, Purky\v{n}ova 123, 612 00, Brno, Czech Republic}
\email{michal.horak2@ceitec.vutbr.cz}
\author{Peter Kepič}
\affiliation[CEITEC]
{Brno University of Technology, Central European Institute of Technology, Purky\v{n}ova 123, 612 00, Brno, Czech Republic}
\author{Jiří Kabát}
\affiliation[CEITEC]
{Brno University of Technology, Central European Institute of Technology, Purky\v{n}ova 123, 612 00, Brno, Czech Republic}
\alsoaffiliation[UFI]
{Brno University of Technology, Faculty of Mechanical Engineering, Institute of Physical Engineering, Technická 2, 616 69, Brno, Czech Republic}
\author{Martin Hájek}
\affiliation[UFI]
{Brno University of Technology, Faculty of Mechanical Engineering, Institute of Physical Engineering, Technická 2, 616 69, Brno, Czech Republic}
\author{Filip Ligmajer}
\affiliation[CEITEC]
{Brno University of Technology, Central European Institute of Technology, Purky\v{n}ova 123, 612 00, Brno, Czech Republic}
\alsoaffiliation[UFI]
{Brno University of Technology, Faculty of Mechanical Engineering, Institute of Physical Engineering, Technická 2, 616 69, Brno, Czech Republic}
\author{Andrea Konečná}
\affiliation[CEITEC]
{Brno University of Technology, Central European Institute of Technology, Purky\v{n}ova 123, 612 00, Brno, Czech Republic}
\alsoaffiliation[UFI]
{Brno University of Technology, Faculty of Mechanical Engineering, Institute of Physical Engineering, Technická 2, 616 69, Brno, Czech Republic}
\author{Tomáš Šikola}
\affiliation[CEITEC]
{Brno University of Technology, Central European Institute of Technology, Purky\v{n}ova 123, 612 00, Brno, Czech Republic}
\alsoaffiliation[UFI]
{Brno University of Technology, Faculty of Mechanical Engineering, Institute of Physical Engineering, Technická 2, 616 69, Brno, Czech Republic}
\author{Vlastimil Křápek}
\email{vlastimil.krapek@ceitec.vutbr.cz}
\affiliation[CEITEC]
{Brno University of Technology, Central European Institute of Technology, Purky\v{n}ova 123, 612 00, Brno, Czech Republic}
\alsoaffiliation[UFI]
{Brno University of Technology, Faculty of Mechanical Engineering, Institute of Physical Engineering, Technická 2, 616 69, Brno, Czech Republic}

\title[]
  {Efficient nanoscale imaging of solid-state phase transitions by transmission electron microscopy demonstrated on vanadium dioxide nanoparticles}

\abbreviations{}
\keywords{}

\begin{document}

\begin{abstract}
We present annular dark field scanning transmission electron microscopy (ADF-STEM) as an efficient, fast, and non-destructive nanoscale tool for monitoring solid-state phase transition. Using metal-insulator transition in vanadium dioxide nanoparticles as an example, we characterize lattice and electronic signatures of the phase transition using analytical transmission electron microscopy including diffraction and electron energy-loss spectroscopy. We demonstrate that ADF-STEM shows a clear contrast across the transition, interpreted with the help of convergent electron beam diffraction as stemming from the crystal-lattice modification accompanying the transition.
In addition, ADF-STEM utilizes 3--6 orders of magnitude lower electron dose when compared to electron microscopy techniques able to reveal the phase transition with the same spatial resolution and universality.
The benefits of ADF-STEM are emphasized by recording a full hysteresis loop for the metal-insulator transition of a single vanadium dioxide nanoparticle. Our study opens the prospect for fast, non-destructive, large-area and nanoscale characterization of solid-state phase transitions.
\end{abstract}

\section{Introduction}

Polymorphic materials, which form more than one stable solid phase, are of both fundamental and technological interest because the associated solid--solid phase transitions (SSPTs) can be used for external control of electronic, magnetic and thermal properties. Metals \cite{Park2009, Kim2004, Chipman1972, Tsutsui2019} and binary metal oxides \cite{Morin1959, Yang2011} represent notable examples of such materials. Besides them, a lot of effort has been recently invested in studying phase transitions of transition metal dichalcogenides (TMDs) \cite{Lin2014, Gautam2024, Lee2023, Song2023, Danz2021}, and low-dimensional structures of other materials \cite{Jacobs2001, Zeng2017, Deng2019, Aryana2021, Han2023, Luo2023}. Especially in these low-dimensional structures, SSPTs and the resulting change of the material properties can be strongly influenced by the quantum size effect or by local inhomogeneities (grain boundaries, defects) whose effect would be averaged out in macroscopic samples. Experimental techniques providing suitable spatial resolution are therefore needed to visualize and fully understand the transition mechanisms. Scanning probe microscopy techniques \cite{Deng2019, Gautam2024} provide high-resolution topography and electronic properties, but these techniques are inherently slow. Electron microscopy techniques, on the other hand, with their fast acquisition and wide range of available signals, are perfect candidates for identifying distinct phases and monitoring the phase transition down to the nanometer or even atomic scale \cite{Lin2021}. The most popular techniques for identifying SSPTs are bright-field (BF) and dark-field (DF) transmission electron microscopy (TEM), which use image contrast changes to observe the phase transition \cite{Jacobs2001, Kim2004, Aryana2021, Lee2023, Song2023, Luo2023}.  Complementary to these two modalities is selected area electron diffraction (SAED), where most SSPTs manifest themselves as differences in diffraction patterns. Due to low-dose illumination and clear imaging of crystal-structure transformations, DF imaging and SAED have also been successfully applied in ultrafast TEM to resolve SSPTs with sub-ps temporal resolution \cite{Park2009, Danz2021, Grinolds2006}. However, to see a contrast change upon SSPT in BF or DF-TEM, the crystals must be specifically oriented, limiting the applicability to polycrystalline samples. In the case of SAED, observation of the SSPT in multiple grains smaller than the aperture size (typically several hundreds of nm) is even nearly impossible due to the overlap of diffraction patterns.

These limitations can be overcome using a scanning TEM (STEM) with a probe focused down to sub-angstrom size. Compared to TEM based on parallel beam illumination, STEM can provide truly atomically-resolved information. Moreover, STEM images allow for more straightforward interpretation than TEM, which often requires numerical image modeling \cite{Kirkland2010}. Furthermore, STEM can be combined with electron energy-loss spectroscopy (EELS), which enables analysis of local elemental composition, bonding environment, or electronic structure \cite{Egerton2011, Colliex2019, Hage2020, Lin2021, Yang20222}. In the context of SSPTs, STEM-EELS could thus elucidate the physics behind the distinct phases of arbitrary-oriented structures. On the other hand, high electron doses required for STEM-EELS measurements can significantly affect the studied materials or even trigger the phase transitions \cite{Zhang2017, Gautam2024}. The type of image contrast in STEM varies with the angular range of the detected scattered electrons. High-angle scattering (typically above \SI{60}{mrad}) provides contrast related to the atomic number. On the other hand, electrons scattered to small angles and collected either at a BF detector or (low-angle) annular DF (ADF) detector are highly influenced by diffraction. The interpretation of BF and ADF STEM images thus often requires simulations. 

In this manuscript, we reveal the significant potential of ADF-STEM for the investigation of SSPT in low-dimensional structures. As a proof-of-principle study, we investigate SSPT of vanadium dioxide (\ce{VO2}), which represents one of the most utilized materials in tunable nanophotonics \cite{Cueff2020}. \ce{VO2} offers ultrafast insulator--metal (monoclinic--rutile) SSPT around the modest transition temperature of \SI{67}{\celsius} \cite{Parra2021, Kepic2021, Jung2022}. In this study, we apply high-resolution TEM imaging, SAED, DF-TEM, ADF-STEM, convergent beam electron diffraction (CBED), and EELS to relate changes in DF-TEM and ADF-STEM images to the insulator--metal (monoclinic--rutile) transition identified by other techniques. Although each of these techniques has been used on its own in previous studies \cite{Felde1997, Sohn2007, Guo2011, Liu2014, Zhang2017, Wu2022, Kim2023, Moatti2019, Krpensky2024}, a comprehensive and correlative investigation of their advantages and disadvantages in the context of SSPT is missing. Besides recording the full SSPT hysteresis of the individual NP by DF-TEM and ADF-STEM, we quantify and compare the electron dose exposure from ADF-STEM and other techniques and demonstrate significant advantages of ADF-STEM.
Our work sets the grounds for the efficient characterization of phase transitions in nanostructured materials using STEM.

\section{Results}

We first inspect the two stable phases of \ce{VO2}, the low-temperature insulating monoclinic phase and the high-temperature metallic rutile phase, with various TEM and STEM techniques. HRTEM imaging and SAED provide direct discrimination of both phases by determining their characteristic crystal lattice. Similarly, the presence or absence of the characteristic plasmon peak in the low-loss EELS directly indicates the presence or absence of free (conducting) electrons specific for the metallic or the insulating phase, respectively. The differences in the core-loss EELS also reflect the differences in the electronic structure of both phases, although their interpretation is less straightforward and relies on the DFT simulations and experimental studies of bulk material. All these techniques utilize high-electron doses (except for SAED, which, on the other hand, exhibits inferior spatial resolution), can be harmful to the sample, and are impractical for intense characterization such as monitoring the hysteresis loop. Interestingly, we observe the contrast between both phases also in the low-dose techniques, namely, DF-TEM and ADF-STEM. Although the interpretation of this contrast is not straightforward, we track its origin to differences in the diffraction patterns of both phases, as evidenced by CBED and SAED. With this, we establish the ability of any of the discussed TEM and STEM techniques to monitor the SSPT in nanostructured materials. Next,  we utilize both low-dose DF imaging methods to study the full hysteresis of the individual \ce{VO2} NP.

\begin{figure}
  \includegraphics{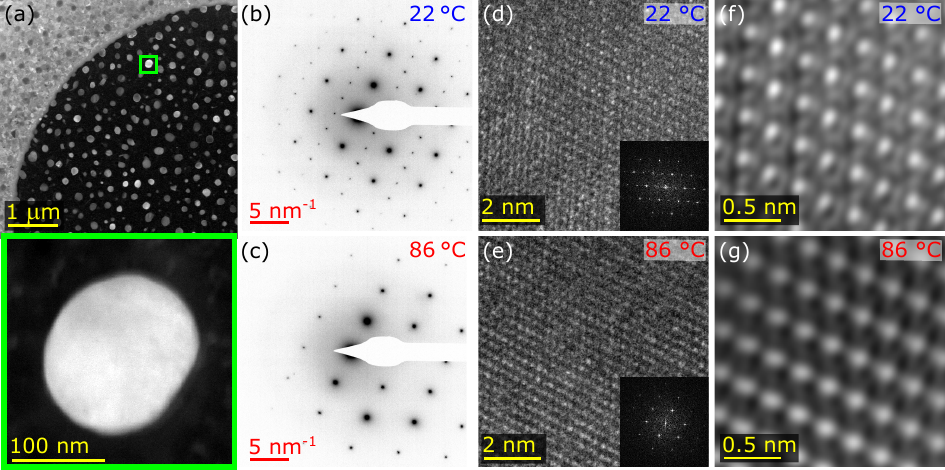}
  \caption{Crystallography of insulating (monoclinic, low-temperature) and the metallic (rutile, high-temperature) phase of the \ce{VO2} nanoparticle. (a) ADF-STEM image of the sample with vanadium dioxide nanoparticles. The inset shows the ADF-STEM image of the selected \ce{VO2} NP. (b,c) SAED patterns recorded at \SI{22}{\celsius} (b) corresponding to the monoclinic lattice with the $[0\,1\,1]$ orientation, and at \SI{86}{\celsius} (c) corresponding to the rutile lattice with the $[-1\,1\,1]$ orientation. The indexation of diffraction spots is available in Fig. S1. (d,e) High-resolution TEM images at \SI{22}{\celsius} (d) and \SI{86}{\celsius} (e). The insets show the 2D FFT of the micrographs. (f,g) Fourier-filtered high-resolution TEM images at \SI{22}{\celsius} (f) and \SI{86}{\celsius} (g)\stxiv{ with overlayed crystallographic model of $[0\,1\,1]$ orientation of the monoclinic phase of \ce{VO2} and the $[-1\,1\,1]$ orientation of the rutile phase of \ce{VO2}}{}.}
  \label{Fig1}
\end{figure}

The full TEM and STEM analysis was performed for the same individual \ce{VO2} NP shown in Fig.~\ref{Fig1}a. For the characterization of the low-temperature insulating monoclinic phase and the high-temperature metallic tetragonal (rutile) phase we utilized the temperatures of \SI{22}{\celsius} and \SI{86}{\celsius}, respectively. The latter temperature was chosen well above the bulk transition temperature to ensure that the transition to the metallic state was complete even in the NP.

The crystal lattice of both phases is characterized with SAED patterns (Fig.~\ref{Fig1}b,c), and high-resolution TEM images (Fig.~\ref{Fig1}d,e) which are further zoomed-in and Fourier-filtered to enhance the visibility of individual atomic columns (Fig.~\ref{Fig1}f,g). Both the SAED and Fourier-filtered HRTEM images indicate the $[0\,1\,1]$ orientation of the monoclinic phase and the $[-1\,1\,1]$ orientation of the rutile phase of this nanoparticle, respectively.

\begin{figure}
  \includegraphics{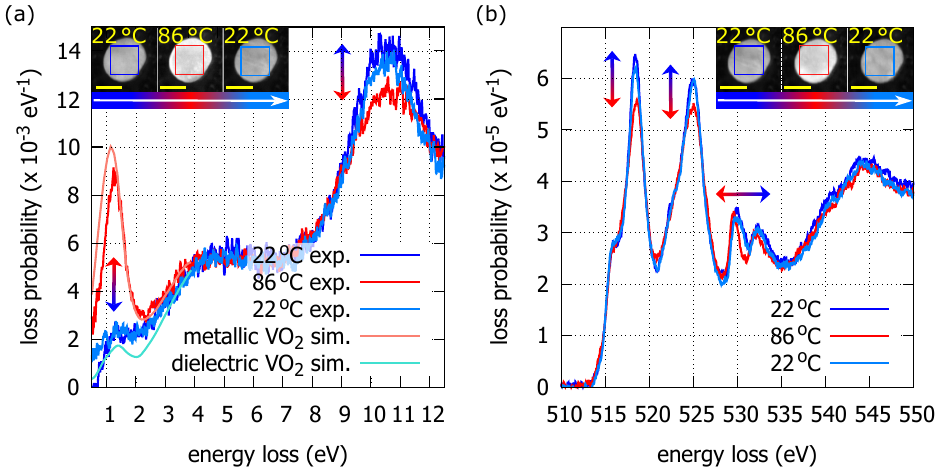}
  \caption{Low-loss (a) and core-loss (b) EELS of the insulator-metal-insulator transition in the \ce{VO2} nanoparticle. (a) Measured EEL spectra of the low-loss region at these three states indicate the insulator-metal-insulator transition by the presence of plasmon peak around \SI{1.2}{\eV} in the metallic phase (\SI{86}{\celsius}, red line) which is absent in the dielectric phases (\SI{22}{\celsius}, blue lines). The phase transition is further confirmed by a very good agreement with calculated low-loss EEL spectra for the metallic (salmon line) and the dielectric (turquoise line) phase. (b) Measured EEL spectra of vanadium L$_{2,3}$--edge and oxygen K--edge at these three states indicate the insulator-metal-insulator transition by the intensity variation of vanadium L$_3$ (at \SI{518.5}{\eV}) and L$_2$ (at \SI{525}{\eV}) peaks and a discernible energy shift within the fine structure of the oxygen K--edge (spectral region from \SI{529}{\eV} to \SI{534}{\eV}) between the dielectric phases (\SI{22}{\celsius}, blue lines) and the metallic phase (\SI{86}{\celsius}, red line). The insets show the ADF-STEM images of the \ce{VO2} nanoparticle. The scale bar in the insets has the length of~\SI{100}{\nano\metre}.}
  \label{Fig2}
\end{figure}

Low-loss energy region characterizes the optical properties of probed material so the low-loss EELS directly distinguishes between the dielectric and the metallic phase of \ce{VO2}. We measured one insulator-metal-insulator transition cycle in the nanoparticle induced by thermal heating of the sample from \SI{22}{\celsius} to \SI{86}{\celsius} followed by cooling back from \SI{86}{\celsius} to \SI{22}{\celsius}. Fig.~\ref{Fig2}a shows the measured and background subtracted low-loss EEL spectra at these three states. The proof of the insulator-metal-insulator transition is the presence of an intense plasmon peak around \SI{1.2}{\eV} in the metallic phase (\SI{86}{\celsius}, red line), which is absent in the dielectric phases (\SI{22}{\celsius}, blue lines). The measured spectra are in excellent agreement with numerical simulations of the EEL spectra calculated for the metallic (salmon line) and the dielectric (turquoise line) phase of the \ce{VO2} nanoparticle. The second peak around \SI{10.5}{\eV} changes its intensity with the phase change and has a 15\% lower loss probability in the metallic phase than in the dielectric phase. Moreover, a very good overlap of both blue lines representing the initial and the final dielectric phase indicates that the phase change is reversible.

Core-loss energy region contains ionization edges of chemical elements within the local (bonding) environment. In the case of \ce{VO2}, the energy region of interest is from \SI{510}{\eV} to \SI{550}{\eV} and contains vanadium L$_3$--edge at \SI{518.5}{\eV} (transitions from $2p_{3/2}$ to unoccupied $3d$ states of V), vanadium L$_2$--edge at \SI{525}{\eV} (transitions from $2p_{1/2}$ to unoccupied $3d$ states of V), and oxygen K--edge with two peaks in the spectral region from \SI{529}{\eV} to \SI{534}{\eV} (transitions of O electrons from $1s$ to hybridized O $2p$ -- V $3d$ states, which are of $t_{2g}$ and $e_g$ symmetry) and a broad peak starting at \SI{535}{\eV} with a maximum around \SI{544}{\eV} (transitions of O electrons from $1s$ to O $2p$ and V $4s$ mixed states) \cite{Hébert2002, Gloter2001, Krpensky2024}. Note that the broad oxygen peak can also be influenced by the presence of unbound oxygen or other oxygen compounds due to contamination. The dielectric and the metallic phase of \ce{VO2} can be distinguished from the fine structure of oxygen K--edge in the spectral region from \SI{529}{\eV} to \SI{534}{\eV}. Fig.~\ref{Fig2}b shows the measured and background subtracted core-loss EEL spectra of the sample at three applied temperatures. The insulator-metal-insulator transition has the following signatures: (i) The loss probability of vanadium L$_3$--edge at \SI{518.5}{\eV} is by 12\% lower and that of L$_2$--edge at \SI{525}{\eV} is by 8\% lower for the metallic phase (\SI{86}{\celsius}, red line) than for the dielectric phases (\SI{22}{\celsius}, blue lines). (ii) There is a discernible energy shift within the fine structure of the oxygen K--edge (spectral region from \SI{529}{\eV} to \SI{534}{\eV}) between the dielectric phases (\SI{22}{\celsius}, blue lines) and the metallic phase (\SI{86}{\celsius}, red line) with the latter shifted about \SI{0.5}{\eV} to lower energies. (iii) Again, a very good overlap of both blue lines representing the initial and the final dielectric phase indicates that the phase change is reversible.

\begin{figure}
  \includegraphics{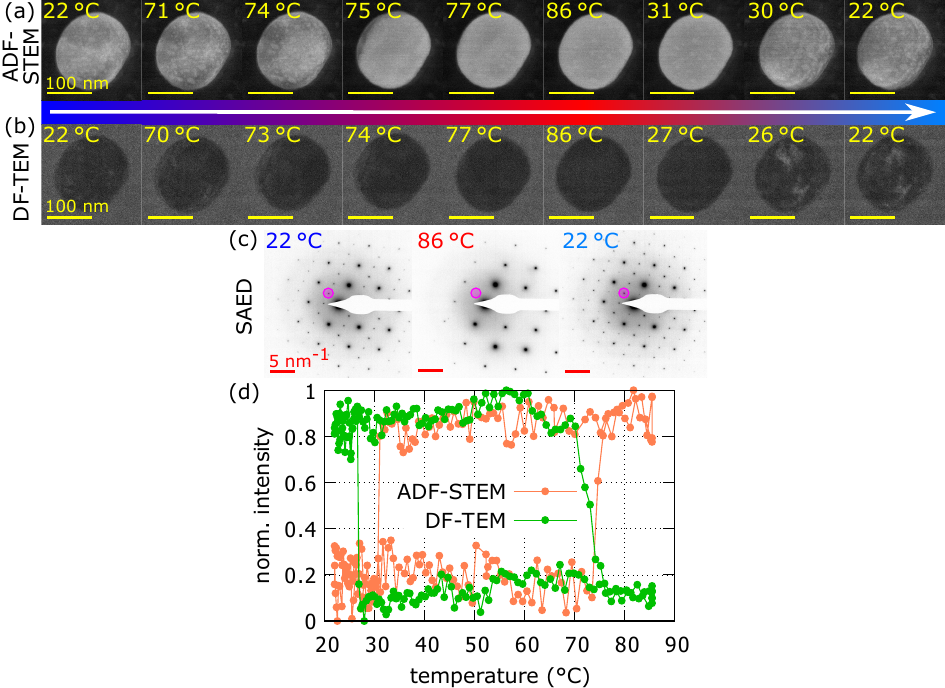}
  \caption{Full evolution of the insulator-metal-insulator transition in the \ce{VO2} nanoparticle. (a) ADF-STEM images showing the insulator-metal transition (gradual change starts at \SI{74}{\celsius} and finishes at \SI{77}{\celsius}) and the metal-insulator transition (fast change between \SI{31}{\celsius} and \SI{30}{\celsius}). (a) DF-TEM images showing the insulator-metal transition (gradual change starts at \SI{73}{\celsius} and finishes at \SI{77}{\celsius}) and the metal-insulator transition (fast change between \SI{27}{\celsius} and \SI{26}{\celsius}). (c) \stxiv{CBED of 3 states (at \SI{22}{\celsius}, at \SI{86}{\celsius}, and back at \SI{22}{\celsius}). The pink area shows the position of the ADF-STEM detector used in (a). (d)}{} SAED of 3 states (at \SI{22}{\celsius}, at \SI{86}{\celsius}, and back at \SI{22}{\celsius}). The pink area shows the position of the objective aperture used in the DF-TEM in (b) situated around the $-1\,0\,0$ diffraction spot of the dielectric phase of the \ce{VO2} nanoparticle. \stxiv{(e)}{(d)} Hysteresis loops extracted from the ADF-STEM (orange) and DF-TEM (green) video. Note that there are minor differences between the initial and final low-temperature micrographs and diffraction patterns due to a slightly different crystallographic orientation of the NP after both (heating and cooling) lattice transformations.}
  \label{Fig3}
\end{figure}

Contrary to previously considered TEM and STEM techniques (HRTEM, SAED, low-loss and core-loss EELS), dark-field imaging (DF-TEM and ADF-STEM) is not expected to provide direct insight into SSPT in general, and the MIT in \ce{VO2} NP in particular. However, we will demonstrate that the imaging contrast varies across the MIT. We will explain that the variation in the DF contrast originates in the transformation of the crystal lattice related to the changes in the diffraction pattern, and we will exploit DF imaging as the fast and low-electron-dose method to monitor the full hysteresis loop of the MIT in \ce{VO2} NP.

To this end, we performed two complete insulator-metal-insulator transition cycles in the nanoparticle induced by thermal heating of the sample from \SI{22}{\celsius} to \SI{86}{\celsius} followed by cooling back from \SI{86}{\celsius} to \SI{22}{\celsius}. We recorded an ADF-STEM video during the first cycle and a DF-TEM video during the second cycle. Both videos are available in the Supplementary information. Fig.~\ref{Fig3}a,b shows selected images from these experiments. ADF-STEM images in Fig.~\ref{Fig3}a show the insulator-metal transition as a gradual change that starts at \SI{74}{\celsius} and finishes at \SI{77}{\celsius}. The metal-insulator transition happens more abruptly between \SI{31}{\celsius} and \SI{30}{\celsius}. Both these transitions are visible as a variation in the contrast and the microstructure in ADF-STEM images caused by the change of crystallography on the nanoparticle during the phase transition. The insulating phase is, in the ADF-STEM signal, darker than the metallic phase. \stxiv{This observation is further confirmed by CBED patterns recorded at three limit temperatures reading \SI{22}{\celsius}, \SI{86}{\celsius}, and back at \SI{22}{\celsius} (Fig.~\ref{Fig3}c) where the pink area shows the position of the ADF-STEM detector.}{} 

DF-TEM images in Fig.~\ref{Fig3}b show the insulator-metal transition as a gradual change starting at \SI{73}{\celsius} and finishing at \SI{77}{\celsius} and the metal-insulator transition as a fast change between \SI{27}{\celsius} and \SI{26}{\celsius}. Both these transitions are visible as a change in the contrast and the microstructure in DF-TEM while the brighter phase is the dielectric one as the DF-TEM is recorded using a diffraction spot which is characteristic for the dielectric phase and not present in the case of the metallic phase. The position of the objective aperture used in the DF-TEM is marked by a pink color in Fig.~\ref{Fig3}d where the SAED patterns recorded at three limit temperatures are shown. Note that there are minor differences between the initial and final low-temperature micrographs and diffraction patterns due to a slightly different crystallographic orientation of the nanoparticle. Fig.~\ref{Fig3}e presents the hysteresis loops extracted from the ADF-STEM and DF-TEM video showing the normalized intensity of ADF-STEM and DF-TEM images of the nanoparticle, respectively. In the case of ADF-STEM, the normalized intensity is higher for the metallic phase of \ce{VO2} (see also Fig.~\ref{Fig3}a), whereas in the case of DF-TEM, the normalized intensity is higher for the dielectric phase of \ce{VO2} (see also Fig.~\ref{Fig3}b). As both hysteresis loops overlap with a minor difference of \SI{4}{\celsius} in the metal-insulator transition the phase change cycle is repeatable within the experimental error.

All the TEM and STEM techniques applicable in the characterization of SSPT have their advantages and disadvantages. They differ in the characterization time, electron dose (and related destructivity to the sample), sensitivity to a specific physical quantity varied during the phase transformation, and the complexity of their interpretation.

SAED, HRTEM, and in an indirect way (through changes in the contrast) also DF imaging (both DF-TEM and ADF-STEM) are sensitive to the changes in the crystal lattice, while EELS is sensitive to the changes in the electronic structure, and low-loss EELS is specifically sensitive to the density of free charge carriers. Regarding the ability to identify individual phases, SAED can identify the phase based on the crystal lattice reconstructed from the observed diffraction pattern and low-loss EELS provides nearly direct identification of phases upon metal-insulator transition through the presence or absence of the plasmon peak. HRTEM and core-loss EELS allow for phase identification after their interpretation based on detailed theoretical modeling; the unprocessed data can be utilized to identify the phase transition but not the individual phases. Finally, DF imaging does not provide sufficient information for phase identification, but the change in the contrast marks the phase transition.

The acquisition times and electron doses of the utilized techniques are provided in Tables~\ref{stab1} and~\ref{stab2}.
Clearly, ADF-STEM imaging stands out among other techniques due to its low electron dose. DF-TEM and SAED are comparable in terms of the electron dose. However, ADF-STEM is superior to SAED in terms of spatial resolution and DF-TEM in terms of universality (as DF-TEM requires a specific orientation of the sample). When compared to the techniques of comparable spatial resolution and universality (EELS; HRTEM offers good spatial resolution but like DF-TEM it requires a specific orientation), ADF-STEM requires the electron dose smaller by 3--6 orders of magnitude.
It therefore offers the possibility to characterize SSPTs with a unique and unprecedented combination of low acquisition time, wide characterization area, nanoscale resolution, and low damage imposed on the sample. The full potential of ADF-STEM would be unleashed in combination with techniques providing detailed physical insight into SSPTs (EELS, SAED). The precharacterization with ADF-STEM allows to identify the transition temperatures (or other thermodynamic parameters) of the forward and backward phase transition, as well as other parameters of the hysteresis loop and their spatial distribution across the sample. This knowledge can be then exploited to perform targeted analysis with more powerful (but also slower and more damaging) techniques instead of tedious spatial and thermal scanning.

\section{Conclusion}

We have presented a novel concept for the experimental characterization of solid-state phase transitions with transmission electron microscopy. It builds upon ADF-STEM imaging as a rapid low-damage method that allows, in a pre-characterization step, to perform multiparametric analysis (typically, spatial and temperature dependences) of wide-area samples and identify the parameter intervals of interest (i.e., regions and temperature intervals), where the phase change occurs. This allows to perform a full analysis, exploiting powerful but demanding methods of electron spectroscopy and diffraction, in a targeted way instead of scanning full parameter space, saving the experimental effort and time and significantly reducing the electron-beam exposition of the sample and related contamination and damage.

In a proof-of-concept experiment, we analyze the metal-insulator transition in \ce{VO2} nanoparticles. We analyze both the metallic and insulating phases, observing differences in conductivity (LL EELS), electronic structure (CL EELS), and crystal lattice (SAED, HRTEM). The contrast between the two phases observed in ADF-STEM is attributed to different diffraction patterns as identified by CBED. Next, we utilize ADF-STEM to characterize a full hysteresis loop for a single \ce{VO2} nanoparticle, observing the insulator-to-metal transition upon heating as a gradual process between \SI{74}{\celsius} and \SI{77}{\celsius}, and the metal-to-insulator transition upon cooling as a rather abrupt process between \SI{31}{\celsius} and \SI{30}{\celsius}. We stress that the transition temperatures and temperature widths can be identified solely by the ADF-STEM imaging, without the need for spectroscopy or diffraction.

ADF-STEM is found to impose the electron dose by 3--6 orders of magnitude lower than the other TEM or STEM methods capable of characterizing the phase transition with comparable spatial resolution and universality. This represents a significant improvement in comparison to the state of the art and opens a new era for the application of transmission electron microscopy in the analysis of solid-state phase transitions.

\section{Methods}

\subsection{Sample fabrication}
Vanadium dioxide nanoparticles were fabricated on a 30-nm-thick silicon nitride membrane of the heating chip by Protochips in a two-step process. First, a 30-nm-thick amorphous film was fabricated by TSST pulsed laser deposition (PLD) system using the following parameters: \SI{248}{\nano\metre} KrF laser, \SI{2}{\joule\per\square\centi\metre}, \SI{10}{\hertz}, 50000 pulses, vanadium target (99.9\% purity, Mateck GmbH), \SI{50}{\milli\metre} substrate-target distance, room temperature, and 5\,mTorr oxygen pressure. Second, the amorphous film was annealed ex-situ for 30 minutes in a vacuum furnace (Clasic CZ Ltd.) at \SI{700}{\celsius} under 15\,sccm oxygen flow to recrystallize into nanoparticles.

\subsection{Transmission electron microscopy}
TEM analysis was performed on TEM FEI Titan equipped with a monochromator, GIF Quantum spectrometer for EELS, and in-situ Fusion Select system by Protochips for heating experiments. Primary beam energy was set to \SI{120}{\kilo\eV} in all experiments to achieve an optimal signal-to-background ratio \cite{Horak2020} and to suppress relativistic losses like the Čerenkov radiation \cite{Horak2015} in the low-loss EELS. The convergence semi-angle in STEM was set to \SI{8.14}{\milli\radian}. The ADF-STEM signal was collected by a Gatan ADF detector capturing electrons scattered into the angle from \SI{16.7}{\milli\radian} to \SI{38.3}{\milli\radian}. The collection angle for both the low-loss and the core-loss EELS was set to \SI{8.3}{\milli\radian}. The DF-TEM images were recorded using a \SI{10}{\micro\metre} objective aperture situated at the position of the $(-1\,0\,0)$ reflection of the dielectric phase of \ce{VO2}. After recording the ADF-STEM and DF-TEM videos using Axon Synchronicity software by Protochips a sequence of images was extracted and processed using the ImageJ software \cite{Schneider2012} and StackReg plugin \cite{Thevanaz1998}. Specifically, we aligned images using the Stack-Reg plugin, created a mask around the nanoparticle, and measured its mean grey value across all images (temperatures). The crystallographic modeling and the SAED indexation were performed using CrysTBox software \cite{Klinger2017} using the monoclinic and rutile \ce{VO2} structures from Ref. \cite{Planer2021}.

\subsection{Simulations}
EEL spectra of a semi-spherical \ce{VO2} nanoparticle with a diameter of \SI{80}{\nano\metre} situated on a 30-nm-thick silicon nitride membrane were calculated using COMSOL Multiphysics utilizing classical dielectric formalism \cite{Abajo2010} following the procedure reported previously \cite{Konecna2018}. Used dielectric functions of the insulator and metallic phase of \ce{VO2} are shown in Figure \ref{FigS2}, and the membrane was approximated by a dielectric constant equal to 4. The electron velocity was set to 0.587 of the speed of light to match the primary beam energy of \SI{120}{\kilo\eV}. Boundary conditions in the form of perfectly matched layers were imposed, and the simulation domain was kept sufficiently large. 

\begin{acknowledgement}

This work is supported by the Czech Science Foundation (project No.~22-04859S). 
We acknowledge CzechNanoLab Research Infrastructure supported by MEYS CR (project No.~LM2023051) for the financial support of the measurements and sample fabrication at CEITEC Nano Research Infrastructure.

\end{acknowledgement}

\begin{suppinfo}

SAED patterns with indexed diffraction spots, dielectric functions of the insulator and metallic phase of VO$_2$ thin film, comparison of experimental parameters and estimation of electron dose for all presented convergent and parallel beam techniques, ADF-STEM and DF-TEM video of the full evolution during the heating-cooling cycle of the phase change in the VO$_2$ nanoparticle.

\end{suppinfo}

\bibliography{single-particle}

\newpage
\section{Supporting Information}

\renewcommand{\thefigure}{S\arabic{figure}}
\setcounter{figure}{0}
\renewcommand{\thetable}{S\arabic{table}} 

\begin{figure}[h!]
  \includegraphics{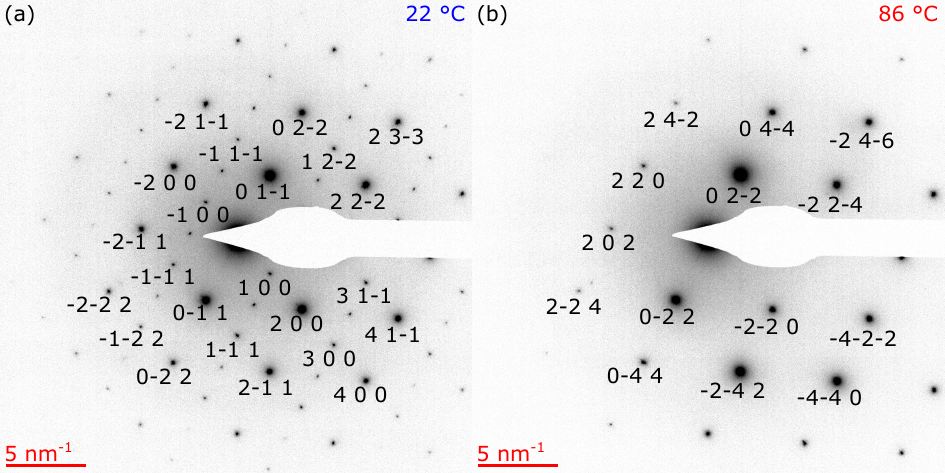}
  \caption{SAED patterns from Fig.~\ref{Fig1}b,c with indexed diffraction spots. SAED patterns were recorded at \SI{22}{\celsius} (a) and \SI{86}{\celsius} (b), and show the $[0\,1\,1]$ orientation of the monoclinic phase and the $[-1\,1\,1]$ orientation of the rutile phase of this \ce{VO2} nanoparticle, respectively.}
  \label{FigS1}
\end{figure}

\begin{figure}[h!]
    \centering
    \includegraphics{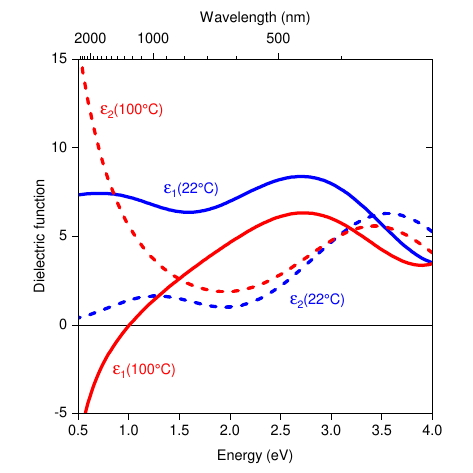}
    \caption{Real (full line) and imaginary (dashed line) parts of the dielectric function of a \SI{30}{nm} VO$_2$ thin film on the silicon substrate obtained at \SI{22}{\celsius} (blue) and \SI{100}{\celsius} (red) in the insulator and metallic phases, respectively.}
    \label{FigS2}
\end{figure}

\clearpage

\begin{table}
    \centering
    \begin{tabular}{|c|c|c|c|c|}
        \hline
        method & exposure time & exposed area & probe current & estimated dose\\\hline
        ADF-STEM & \SI{0.8}{\micro\second} & $\sim$\,\SI{1}{\square\nano\meter} & $\sim$\,\SI{100}{\pico\ampere} & $8\times 10^{-17}$ \SI{}{\coulomb\per\square\nano\meter} \\
        LL-EELS & \SI{0.3}{\milli\second} & $\sim$\,\SI{1}{\square\nano\meter} & $\sim$\,\SI{100}{\pico\ampere}& $3\times 10^{-14}$ \SI{}{\coulomb\per\square\nano\meter}\\
        CL-EELS & \SI{1}{\second} & $\sim$\,\SI{1}{\square\nano\meter} & $\sim$\,\SI{100}{\pico\ampere} & $1\times 10^{-10}$ \SI{}{\coulomb\per\square\nano\meter}\\\hline
    \end{tabular}
    \caption{\label{stab1}Experimental parameters and estimated electron dose for convergent beam techniques (ADF-STEM, LL-EELS, and CL-EELS) required to obtain the signal allowing to observe the MIT in \ce{VO2} nanoparticle at 1 pixel. The exposed area corresponds to the cross-section of the focused electron beam which is approximately a circle with a diameter of $\sim$\,\SI{1}{\nano\meter}.}
\end{table}

\begin{table}
    \centering
    \begin{tabular}{|c|c|c|c|c|}
        \hline
        method & exposure time & exposed area & probe current & estimated dose\\\hline
        HRTEM & \SI{1}{\second} & $\sim$\,\SI{0.005}{\square\micro\meter} & $\sim$\,\SI{3}{\nano\ampere} & $6\times 10^{-13}$ \SI{}{\coulomb\per\square\nano\meter}\\
        SAED & \SI{2}{\second} & $\sim$\,\SI{9}{\square\micro\meter} & $\sim$\,\SI{3}{\nano\ampere} & $7\times 10^{-16}$ \SI{}{\coulomb\per\square\nano\meter}\\
        DF-TEM & \SI{1}{\second} & $\sim$\,\SI{9}{\square\micro\meter} & $\sim$\,\SI{3}{\nano\ampere} & $3\times 10^{-16}$ \SI{}{\coulomb\per\square\nano\meter} \\\hline
    \end{tabular}
    \caption{\label{stab2}Experimental parameters and estimated electron dose for parallel beam techniques (HRTEM, SAED, and DF-TEM) required to obtain the micrograph allowing to observe the MIT in \ce{VO2} nanoparticle. The exposed area corresponds to the cross-section of the parallel electron beam which is approximately a circle with a diameter of $\sim$\,\SI{70}{\nano\meter} in the case of HRTEM, and of $\sim$\,\SI{3}{\micro\meter} in the case of SAED and DF-TEM, respectively. \stxiv{Note that the resulting electron dose required for DF-TEM is underestimated as it is calculated with parameters for fast video image and not for parameters to obtain a high-quality image as in the case of SAED and HRTEM.}{We note that the electron dose for DF-TEM is lower than the dose for SAED, since the former refers to the fast image recorded during the video sequence and the latter to the high-quality image with a high signal-to-noise ratio necessary for observation of all the diffraction spots required for the phase identification (including the weak ones). For the same quality of the image, the DF-TEM would require significantly higher dose. However, for our purpose of monitoring the SSPT this is not necessary.}}
\end{table}

\end{document}